\documentclass{sigchi}

\usepackage{balance}  
\usepackage{graphics} 
\usepackage{times}    
\usepackage{url}      

\makeatletter
\def\url@leostyle{%
  \@ifundefined{selectfont}{\def\UrlFont{\sf}}{\def\UrlFont{\small\bf\ttfamily}}}
\makeatother
\def\pprw{8.5in}
\def\pprh{11in}

\usepackage[pdftex]{hyperref}
\hypersetup{
pdftitle={SIGCHI Conference Proceedings Format},
pdfauthor={LaTeX},
pdfkeywords={SIGCHI, proceedings, archival format},
bookmarksnumbered,
pdfstartview={FitH},
colorlinks,
citecolor=black,
filecolor=black,
linkcolor=black,
urlcolor=black,
breaklinks=true,
}

\begin{document}

\title{3DTouch: A wearable 3D input device with an optical sensor and a 9-DOF inertial measurement unit}

\numberofauthors{2}
\author{
  \alignauthor Anh Nguyen\\
    \affaddr{Department of Computer Science}\\
    \affaddr{University of Wyoming}\\
    \email{anguyen8@uwyo.edu}\\
  \alignauthor Amy Banic\\
    \affaddr{Department of Computer Science}\\
    \affaddr{University of Wyoming}\\
    \email{abanic@cs.uwyo.edu}\\
}

\maketitle

\begin{abstract}
We present 3DTouch, a novel 3D wearable input device worn on the fingertip for 3D manipulation tasks. 3DTouch is designed to fill the missing gap of a 3D input device that is self-contained, mobile, and universally working across various 3D platforms. This paper presents a low-cost solution to designing and implementing such a device. Our approach relies on relative positioning technique using an optical laser sensor and a 9-DOF inertial measurement unit. 

3DTouch is self-contained, and designed to universally work on various 3D platforms. The device employs touch input for the benefits of passive haptic feedback, and movement stability. On the other hand, with touch interaction, 3DTouch is conceptually less fatiguing to use over many hours than 3D spatial input devices. We propose a set of 3D interaction techniques including selection, translation, and rotation using 3DTouch. An evaluation also demonstrates the device's tracking accuracy of 1.10 mm and 2.33 degrees for subtle touch interaction in 3D space. Modular solutions like 3DTouch opens up a whole new design space for interaction techniques to further develop on.
\end{abstract}




\section{Introduction}
3D applications appear in every corner of life in the current technology era. Besides traditional computer games, and modeling applications, 3D technology has also powered browsers (e.g., with WebGL), touch devices, home theaters, even large visualization platforms such as the Cave Automatic Virtual Environment (CAVE) and many more. However, there is not yet a universal 3D input device that is self-contained and can be used across different platforms. Recent years have witnessed a wide variety of input devices \cite{3dui2013bowman}. Desktop input devices such as traditional mice, keyboards, or 3D mice (e.g., 3Dconnexion SpaceNavigator) provide stability and accuracy; however, they are not portable for spatial environments such as the CAVE. Mobile touch devices~\cite{nguyen2013low} provide intuitive and direct input, but the working space is limited within the screen area. While voice input is convenient, it is not intuitive for users to give voice commands for performing complex 3D interaction tasks (e.g., rotate the red cube 60 degree around z-axis). Although these devices have their unique advantages, they are usually designed for a single certain platform.

One input method to interact with 3D applications is using 3D mid-air gestures, which are popularized by commodity devices like Kinect and Wiimote. However, serving as a type of 3D input beyond the purpose of entertainment, they are subject to a major drawback of fatigue. In 2011, a study \cite{nancel2011mid} showed that 3D mid-air gestures with bare hands are more tiring than 1D and 2D gestures with hand-held input devices (e.g., smartphones, or remote controls). 

\begin{figure}[!h]
\centering
\includegraphics[width=1.0\columnwidth]{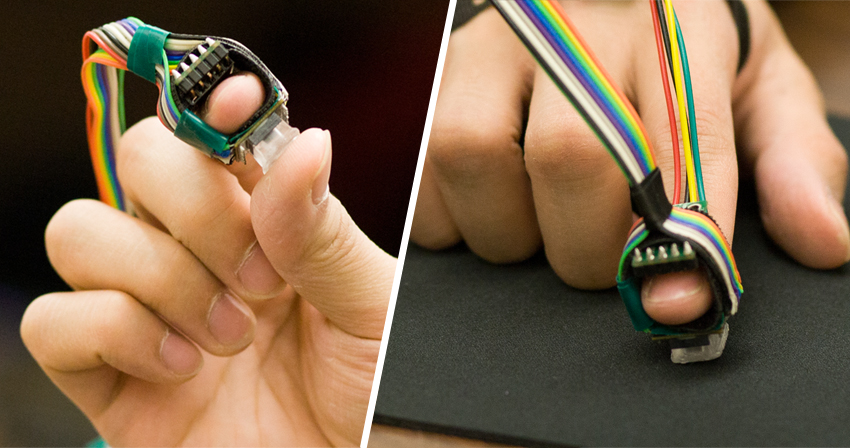}
\caption{3DTouch - a novel 3D input device worn on the fingertip
}
\label{fig:figure_device}
\end{figure}

Touch interaction is another way to interact with 3D applications. Unlike spatial interaction, touch interaction has a subtle neat advantage that users can feel natural passive haptic feedback on the skin via sense of touch. Touch gestures are conceptually less fatiguing than 3D mid-air gestures. Moreover, the touch surface keeps the hand steady and thus increasing the stability and accuracy of finger movements. A variety of creative research works have then brought touch interaction to surfaces that are not inherently touch-sensing capable such as tables \cite{bi2011magic}, walls \cite{matsushita1997holowall}, clothes \cite{saponas2011pockettouch}, skin \cite{harrison2010skinput, harrison2011omnitouch, sato2012touche}, conductive surfaces \cite{sato2012touche} (e.g., the metal door knob, and even liquids), or virtually any flat surface using a combination of a depth-sensing camera and a projector \cite{harrison2011omnitouch, mistry2009sixthsense}. 

In this paper, we present 3DTouch, a thimble-like 3D touch input device worn on the user's fingertip. 3DTouch is self-contained, and universally working on various platforms (e.g., desktop, and CAVE). The device employs touch input for the benefits of passive haptic feedback, and movement stability. On the other hand, with touch interaction, 3DTouch is conceptually less fatiguing to use over many hours than spatial input devices.

3DTouch allows users to perform touch interaction on many surfaces that can be found in an office environment (e.g., mousepad, jeans, wooden desk or paper). When mounted on the tip of index finger, the user can perform touch interaction on the other hand's palm, which serves as the touch pad (Figure~\ref{fig:figure_device}). 3DTouch fuses data reported from a low-resolution, high-speed laser optical sensor, and a 9-DOF inertial measurement unit (IMU) to derive relative position of a pointer in 3D space. The optical sensor, usually found in traditional computer mice, determines the direction and magnitude of movement of the pointer on a virtual 2D plane. And the 9-DOF IMU determines the orientation of the plane. Since we would like to keep the 3DTouch interface simple with no buttons, a gesture recognition engine was developed to allow users to make gestural commands. Based on the data from optical sensor, we used classification techniques to reliably recognize simple gestures such as: tap, double-tap, and press gesture.

The contributions of this paper are:
\begin{enumerate}
\item A novel, low-cost technique to turn a finger or a thumb into 3D touch input device using an optical sensor and a 9-DOF IMU.
\item A set of 3DTouch interaction techniques.
\item An evaluation demonstrating the accuracy of 3DTouch across various surfaces of different materials and shapes.
\end{enumerate}

\section{Related Work}
3DTouch is an interdisciplinary research project that crosses various fields. In this section, we review the related literature in the areas of 3D User Interfaces (3DUI), finger-worn interfaces, and touch interaction.

\subsection{3D User Interfaces}
Motion tracking systems are widely used in 3DUI community because of their capability of sensing position, orientation, and velocity of one or more objects. These 6-DOF position trackers can be based on many different technologies, such as those using electromagnetic fields (e.g., Polhemus Liberty), optical tracking (e.g., NaturalPoint OptiTrack \cite{optiTrack}), or hybrid ultrasonic/inertial tracking (e.g., Intersense IS900). All of these, however, share the limitation that some external fixed reference (e.g., a base station, a camera array, or an emitter grid) must be used. While ultrasonic and electromagnetic tracking techniques are susceptible to environment interference, optical tracking is subject to the inherent problem of occlusion \cite{welch2002motion}.

Inertial tracking systems, on the other hand, can be self-contained and require no external reference. They use technologies such as accelerometers, gyroscopes, and compasses to sense their change in orientation \cite{luinge1999estimating}. While devices equipped with such sensors (e.g., Wiimote, air mice, and smartphones) are capable of serving as a 3D pointing device, they have only been used to translate objects on a fixed 2D plane (e.g., the TV screen). 3DTouch, with 5 degrees of freedom, does not only serve as a 3D pointing device, but also enables users to rotate and translate objects in 3D space.

\subsection{Finger-worn Interfaces}
Early work in instrumenting the human finger was conducted in the 3DUI research community. Ring Mouse \cite{bowman20043d} is a small, ring-like device, with two buttons, worn along the index finger. It uses ultrasonic tracking, but generates only position information. With a similar design to that of Ring Mouse, FingerSleeve uses a 6-DOF magnetic tracker to report position and orientation \cite{zeleznik2002pop}. The drawback of these devices is that they are not self-contained, relying on an external tracking system. 

Using magnetic field sensing techniques, several projects have explored augmenting the finger with a small magnet. With Abracadabra \cite{harrison2009abracadabra}, users wear a magnet on their finger to provide 1D and 2D input to a mobile device. On the other hand, FingerFlux \cite{weiss2011fingerflux} provides simulated haptic feedback to the fingertip when operating above an interactive tabletop. While these devices bring more functionality to the finger, they do not support 3D and always-available input. By mounting a Hall sensor grid on the index fingernail, and a magnet on the thumbnail, FingerPad turns pinched fingertips into a touch pad \cite{chan2013fingerpad}. However, the input space enabled by FingerPad is only 2D. uTrack \cite{chen2013utrack} turns the fingers and thumb into a 3D input device. As a magnet is worn on the thumb, and two magnetometers are worn on the fingers, uTrack is a self-contained 3D input device. However, it is not a full 6-DOF input device and can only serve as a 3D pointing device.

Other researchers explored mounting cameras on the body \cite{harrison2011omnitouch, ni2009disappearing, yang2012magic} for truly ubiquitous use. Logisys's Finger Mouse, a cylinder-shaped optical mouse, brings the traditional mouse control to the finger \cite{logisysFingerMouse}. Extending this concept, Magic Finger \cite{yang2012magic} allows users to recognize 32 different textures for contextual input by augmenting the finger with a high-resolution optical sensor. While these two projects are closely related to 3DTouch in using optical sensors, none of them had the goal of turning the finger into a 3D input device.

\subsection{Extra Dimensions of Touch Interaction}
Many mobile touch devices only utilize the 2D position of a touch contact being made on the surface. However, other auxiliary information of a touch interaction has also proved to be useful such as: the shape \cite{wilson2008bringing, cao2008shapetouch} or size \cite{benko2006precise} of the contact region, the orientation of the finger making contact \cite{wang2009detecting}, and even the touch pressure \cite{ramos2004pressure}. While the size of the contact region was used to improve the precision of selection techniques \cite{benko2006precise}, attributes such as the shape of the contact region \cite{wilson2008bringing, cao2008shapetouch}, orientation of the finger \cite{wang2009detecting}, and touch pressure \cite{ramos2004pressure} were additional inputs for the application to deliver pseudo-haptic feedback to users.

Using a 9-DOF IMU mounted on the fingernail, 3DTouch leverages the finger orientation to augment the 2D input from the optical sensor into 3D input. And the pressure dimension is used to enable \textit{press gesture}, conceptually similar to a mouse-click gesture. Unlike the popular tap gesture on touch devices, press gesture allows the user to make selection commands without lifting finger off the surface, thus reducing workload for the finger joint.

\section{Hardware Prototype}
An open problem of spatial tracking is how to build a 6-DOF system that is self-contained, and capable of tracking its own position and orientation with high levels of accuracy and precision \cite{3dui2013bowman}. With 3DTouch, our approach is to fuse data from a 9-DOF IMU and a laser optical sensor to derive position and orientation.

\subsection{Inertial Measurement Unit}
Pololu MinIMU-9 v2 is a 9-DOF IMU that packs an L3GD20 3-axis gyro, an LSM303DLHC 3-axis accelerometer and 3-axis magnetometer onto a tiny 0.8" x 0.5" board. We selected such an IMU with 9 degrees of freedom because when applying Kalman filter \cite{marins2001extended}, the estimates of orientation would be more precise than those based on a single measurement alone.

\subsection{Optical Flow Sensor}
We used Pixart ADNS-9800 laser optical flow sensor with a modified ADNS-6190-002 lens. The reason we chose a laser sensor is that they work on a larger number of surfaces than LED-based optical sensors. ADNS-9800, often found in modern laser gaming mice, comprises a sensor and a vertical-cavity surface-emitting laser (VCSEL) in a single chip-on-board (COB) package. The sensor is a high resolution (i.e., up to 8200 cpi), black-and-white camera (30 x 30 pixels). However, for the purpose of tracking movements, we manually programmed the resolution down to 400 cpi for a higher frame rate.
This sensor is then wired to an application printed circuit board (PCB) designed according to the schematic diagram in the datasheet \cite{pixartADNS9800}. This PCB streams data from the sensor to an Arduino UNO R3.

\subsection{Physical Form}
\begin{figure}[!h]
\centering
\includegraphics[width=1.0\columnwidth]{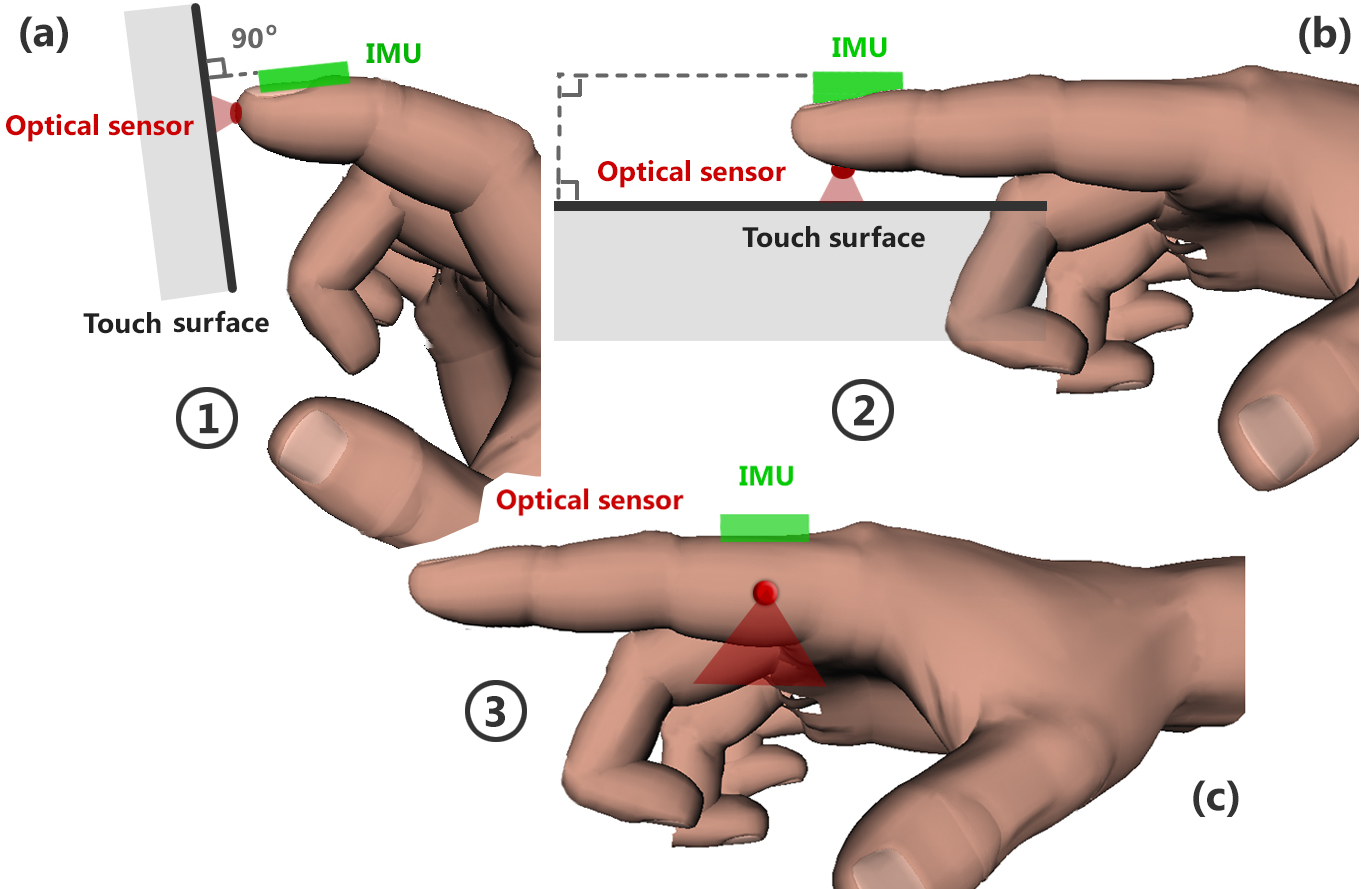}
\caption{Three possible form factors of 3DTouch: (a) Form factor 1: The optical sensor is mounted on the fingertip, below the fingernail. (b) Form factor 2: The optical sensor is placed on the fingerpad. (c) Form factor 3: The finger can serve as a pointing device with 3DTouch worn as a ring. In all three form factors, the IMU is placed on the finger for the finger orientation to be utilized.
}
\label{fig:figure_form}
\end{figure}

The device needs to be small enough to be worn on the user's finger. We mounted the IMU on top of the fingernail so that we can utilize the finger's orientation. Figure~\ref{fig:figure_form} illustrates three possible form factors of 3DTouch. In form factor 1, if small enough, the optical sensor could be mounted on the fingertip. In form factor 2, the optical sensor could be placed on the fingerpad. In form factor 3, 3DTouch, worn as a ring, can be used as a pointing device, and the finger does not have to perform touch interaction with a surface. The third form factor enables users to turn their finger into a pointing device, and use the thumb to perform touch gestures such as tap and double-tap with the optical sensor. Beyond these three proposed form factors, 3DTouch can be usable when being worn on the thumb as well.

Our prototype presented in this paper (Figure~\ref{fig:figure_device}) was implemented according to form factor 2. An user can transform from form factor 2 into form factor 3 by simply pushing 3DTouch further towards the palm. 3DTouch has the shape of a thimble, which is an adjustable Velcro strap used to hold the sensors. The IMU is mounted on top of the fingernail, and the optical sensor is on the fingerpad (Figure~\ref{fig:figure_device}).

\subsection{Computer Interfacing}
The IMU and optical sensor stream data to an Arduino UNO board. The Atmega16U2 microcontroller on Arduino then applies Kalman filtering to the data from the IMU, and synchronizes the orientation result with relative position data from the optical sensor. The fused data are then streamed to a computer, which is an HP ProBook 4530s running Ubuntu 12.04. An USB cable is used to connect Arduino UNO to the computer for evaluation purposes. This wired connection later could be replaced by a wireless solution using a pair of XBee modules.

\subsection{Lens augmentation}
We a thin layer of elastic rubber of 2.0 mm height around curvature of the ADNS-6190-002 lens of the optical sensor. This allows the distance from the lens to the touch surface to be adjustable from 2.4 mm to 4.4 mm by applying pressure. This augmentation enables our gesture recognition engine to sense pressure as well.

\section{Gesture Detection}
For the device to be usable, we decided to implement the basic touch gestures of tap and double-tap. A novel \textit{press gesture} is also proposed. 
This section explains the algorithms used to enable the tap, double-tap, and press gestures.

\subsection{Sensing Contact}
To sense contact with a surface, Magic Finger relies on rapid changes in the pixel contrast level of the sensor image \cite{yang2012magic}. This approach requires continuous reading of the image pixels, and performing the calculation to derive the change in contrast level.
However, we took a simpler, yet effective approach by monitoring the surface quality (SQUAL) values reported directly by the ADNS-9800 sensor board. As described in the datasheet \cite{pixartADNS9800}, SQUAL ranges from 0-169, and becomes nearly zero if there is no surface below the sensor.

However both approaches of using image contrast level, and SQUAL are still optical techniques to sense contact. Hence, the sensing accuracy is affected by such variables as environment lighting condition, surface texture, and the lift detection (Z-height) setting programmed to the optical sensor. Different surfaces will have different lift detection values with the same setting due to different surface characteristic \cite{pixartADNS9800}.

\subsection{Tap Gesture}
The SQUAL, and x-y increments (e.g., X\_DELTA, and Y\_DELTA) values are used to measure tap gesture. A tap gesture is recognized when there is a rapid change, within 300 ms timespan, in SQUAL from 0 to 40, and in x/y movements between +/-5 units. These settings are specific values for mousepad texture only. When using a texture recognition engine based on Support-Vector Machines \cite{yang2012magic}, it is possible to load the correct settings for corresponding textures.

\subsection{Double-Tap Gesture}
Similar to a double-click gesture, we needed to continuously monitor the tap gestures. If two tap gestures take place within a certain pre-defined time span, then a double-tap gesture is fired.
Microsoft Windows 7 sets 500ms as the default time span for a double-click \cite{windowsDoubleClick}. However, this should be an adjustable setting for users, and for the purposes of testing we set it to be 200-500ms. 

On the other hand, for a double-tap gesture to be recognized, two subsequent taps need to take place at the same position. This is difficult to achieve with optical sensing because there is always noise when the sensor is lifted off the surface. After pilot testing 300 double-tap gestures, we defined the offset distance for two subsequent taps to be recognized as a double-tap to be +/-15 for mousepad texture.

\subsection{Press Gesture}
The lift detection distance for ADNS-9800 ranges from 1-5 mm \cite{pixartADNS9800}. As the fixed height of the ADNS-6190-002 is 2.4 mm, the 2.0 mm thin layer of rubber allows the sensor to still recognize the surface within the 2.4 - 4.4 mm range. For the mousepad texture, an average SQUAL value of 40 corresponds to 2.4 mm lift-off distance under normal indoor light condition. We continuously monitor and detect a press gesture when the SQUAL values reach 40 or above.

This gesture reduces workload for the finger joint as users do not have to lift their finger off the surface. However, it is subject to many other environmental factors such as surface texture, and lighting condition. A mechanical push button may be a more reliable alternative.

\section{3DTouch Interaction Techniques}
This section describes how a single 3DTouch device, worn on a finger or thumb, can be used to perform 3D interaction techniques of selection, translation, and rotation. Interaction techniques utilizing more than one piece of 3DTouch are discussed in the Future Work section, and are not within the scope of this paper. The interaction techniques presented in this section are implemented in Virtual Reality User Interface (Vrui) framework \cite{kreylos2008environment}, which allows 3D applications to run on a wide variety of platforms such as desktops, wall displays and CAVEs.

\subsection{Selection}
\begin{figure}[!h]
\centering
\includegraphics[width=1.0\columnwidth]{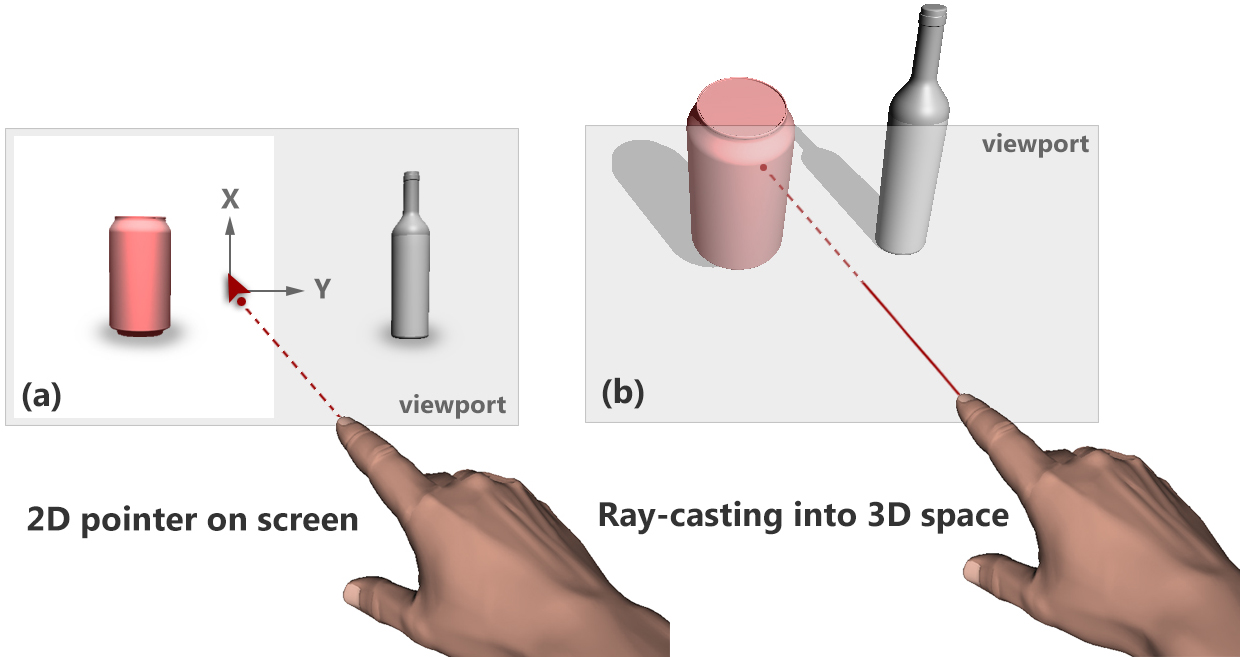}
\caption{(a) Moving the 2D pointer to the left half of the 2D plane to select the soda can. (b) Pointing at the soda can in 3D space to select it.}
\label{fig:figure_selection}
\end{figure}

3DTouch is capable of sensing the absolute 3-DOF orientation of the finger wearing the device. Hence, we propose to use the traditional Ray-Casting technique \cite{bowman20043d} to select an object in 3D space. With ray casting, the user points the finger wearing 3DTouch at objects with a virtual ray that defines the direction of pointing (Figure~\ref{fig:figure_selection}b). More than one object can be intersected by the ray; however, only the one closest to the user should be selected. On a 2D plane such as the TV screen, the user can point up-down and left-right to move the 2D pointer around (Figure~\ref{fig:figure_selection}a).

After a ray is pointed at an object, a tap gesture can be performed to make the selection command. For the selection technique, the form factor 3 is the most suitable because the user can use their finger as a pointing device and give selection commands by performing tap and double-tap gestures. Since 3DTouch does not support absolute positioning, the casted ray always starts from a pre-configured point (e.g., the middle bottom of the screen).

\subsection{Translation}
\begin{figure}[!h]
\centering
\includegraphics[width=1.0\columnwidth]{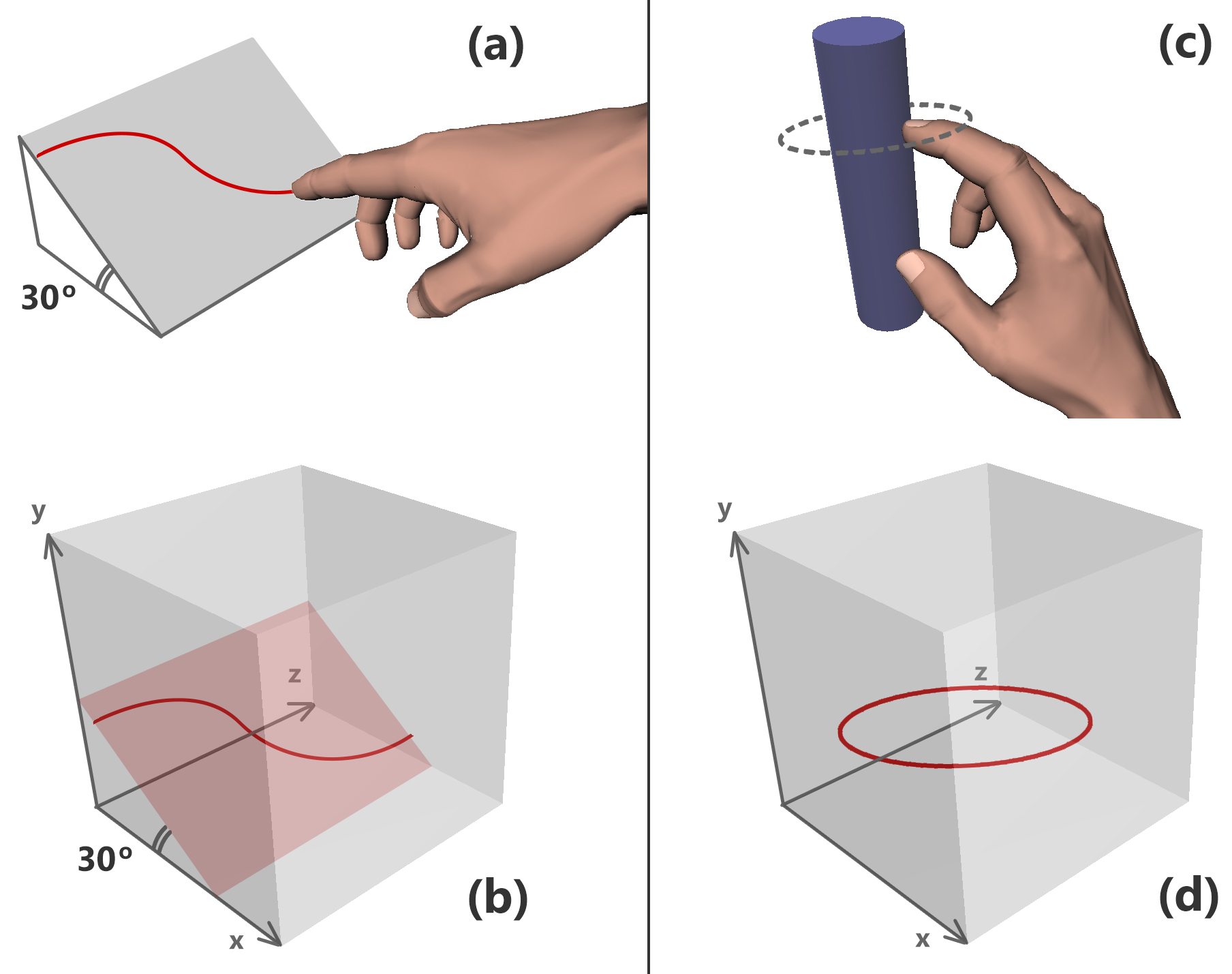}
\caption{(a) The 3DTouch user is drawing a curve (red) on a flat surface, which makes $30^{\circ}$ with the ground. (b) In the 3D VE, a curve is generated on a 2D plane, which also makes $30^{\circ}$ with the XZ plane. (c) The 3DTouch user is touching around the curved surface of a cylinder. (d) In the 3D VE, a circle with diameter proportional to that of the cylinder is generated.
}
\label{fig:figure_translation}
\end{figure}

With an optical sensor, 3DTouch is capable of drawing or translating an object on a 2D plane. However, this plane's orientation is adjustable by the 3-DOF orientation of the user finger. Figure~\ref{fig:figure_translation} illustrates two examples of how the actual touch movements map to a 3D virtual environment (VE). This interaction technique can be applied to both object and screen translation. With 3DTouch, touch interaction can be performed on flat surfaces as well as curved surfaces (see Figure~\ref{fig:figure_translation}c).

\subsubsection{Deriving the orientation of the 2D touch plane}
In form factor 1 (Figure~\ref{fig:figure_form}a), the IMU is perpendicular to the touch surface as illustrated in Figure~\ref{fig:figure_translation}a/c, we added 90 degrees to the \textit{pitch} angle reported from the IMU to achieve the 2D plane orientation.

In form factor 2 (Figure~\ref{fig:figure_form}b), the IMU is parallel with the touch surface, the orientation of the 2D plane is equal to the orientation of the finger.

\subsection{Rotation}
\begin{figure}[!h]
\centering
\includegraphics[width=1.0\columnwidth]{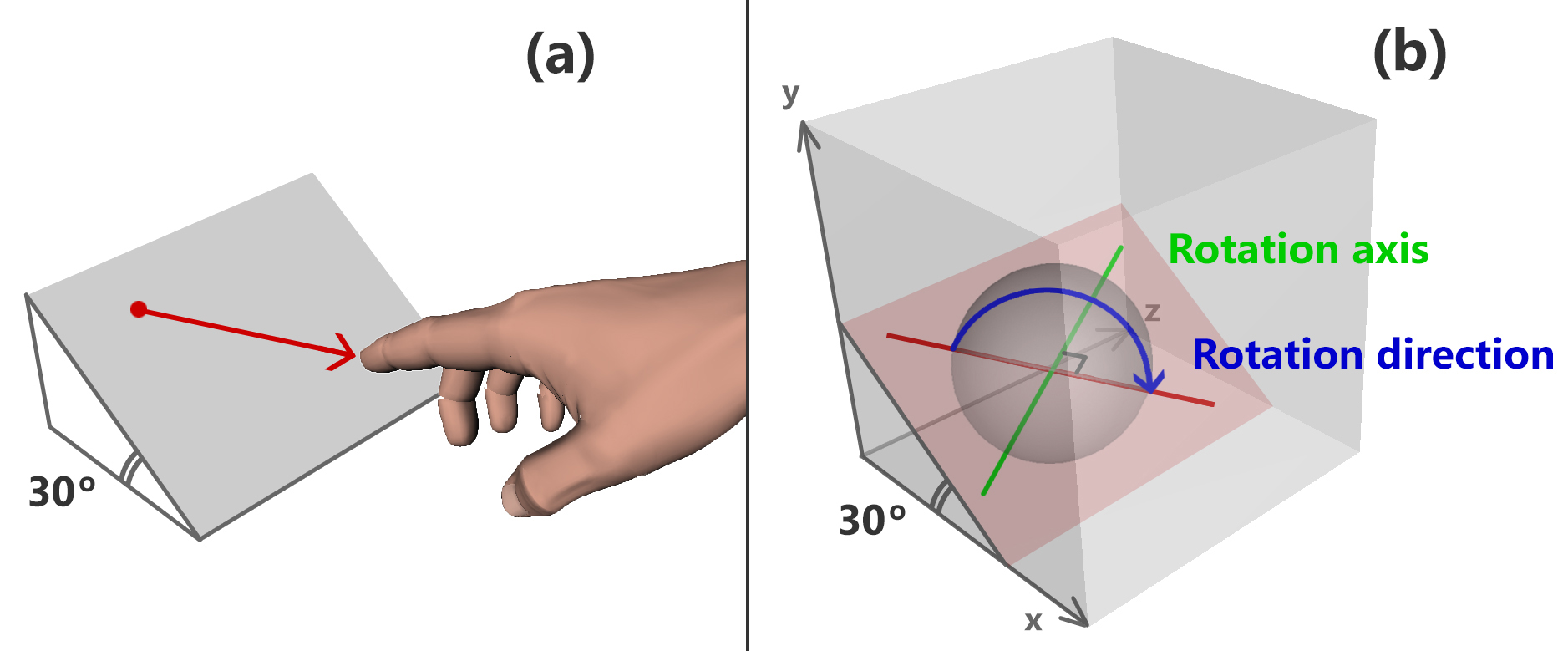}
\caption{(a) The user is drawing a vector (red) on a flat surface, which makes $30^{\circ}$ with the ground. (b) In the 3D VE, the sphere is rotated around the rotation axis by an angle proportional to the vector length. The rotation axis is perpendicular to the vector on the 2D plane, which also makes $30^{\circ}$ with the XZ plane.
}
\label{fig:figure_rotation}
\end{figure}

Similar to translation, the user draws a vector on a surface to rotate a virtual object in focus. And the object will be rotated around the rotation axis, which is perpendicular to the drawn vector on the same 2D plane. The length of the vector drawn is proportional to the rotation angle. And the direction of the vector determines the rotation direction. Figure~\ref{fig:figure_rotation} illustrates an example of how the drawn vector is used to derive the rotation in a 3D VE.

\section{Evaluation of Tracking Accuracy}
We conducted an experiment to evaluate the 3D tracking accuracy of our device with respect to the ground truth across multiple surfaces. We compared the 3D position and 3D orientation reported by 3DTouch, against the data obtained using NaturalPoint OptiTrack motion tracking system \cite{optiTrack}. In this experiment, we assumed the data obtained from the OptiTrack system to be the ground truth. The OptiTrack system reported a maximum mean error below 0.8 mm throughout the whole experiment.

\subsection{Setup}
3DTouch and OptiTrack both streamed their data via wired connections to a Linux machine with a dual-core 2.1GHz CPU with 4GB of RAM. On this machine, a program written in C++ synchronized and logged the samples at 50Hz. The device was configured in the form factor 1, and worn by the first author on the index finger. To capture the movements of 3DTouch, we setup the tracking volume using 12 Flex-13 cameras sampling at 120 Hz. A rigid body, composed of three reflective markers, was mounted on top of 3DTouch. 

\begin{figure}[!h]
\centering
\includegraphics[width=0.8\columnwidth]{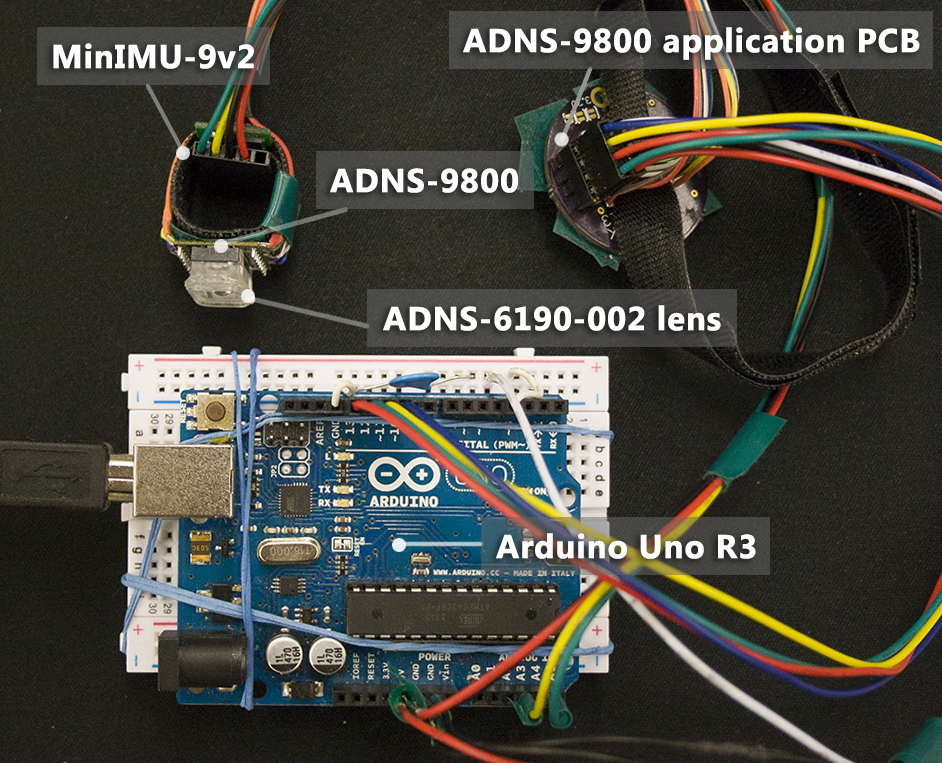}
\caption{The setup of 3DTouch including the Arduino UNO R3, an optical sensor, and IMU, and an sensor application board (purple).
}
\label{fig:figure_setup}
\end{figure}

\subsection{Experimental Design}
\begin{figure}[!h]
\centering
\includegraphics[width=0.8\columnwidth]{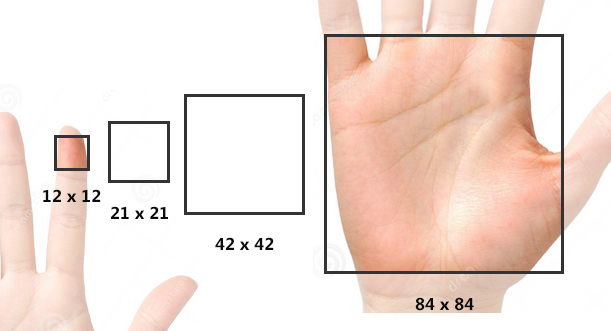}
\caption{We designed 4 different target sizes: 12 x 12mm, 21 x 21mm, 42 x 42mm, and 84 x 84mm. 12 x 12mm is the smallest touch area usable by a previous work FingerPad. The largest area is designed according to the average human palm size, which is the touch area for the target mobile applications of 3DTouch.
}
\label{fig:figure_target_sizes}
\end{figure}

Since the surface texture is the factor affecting the optical sensing accuracy, we tested the device across 3 textures: mousepad, wooden desk, and jeans. These are three of the environmental textures used as contextual input for Magic Finger \cite{yang2012magic}. For each texture, we designed 4 different target sizes: 12 x 12mm, 21 x 21mm, 42 x 42mm, and 84 x 84mm (Figure~\ref{fig:figure_target_sizes}). We chose 12 x 12mm as the smallest size because that is the smallest touch area usable by a previous work \cite{chan2013fingerpad}. The largest area is designed according to the average human palm size \cite{agnihotri2006determination}, which is the touch area for the target mobile applications of 3DTouch. 

For each target size, we performed drawing 6 basic shapes: horizontal line, vertical line, diagonal line, triangle, square, and circle. These basic shapes are the building blocks for users to perform 3D interaction techniques and 2D gestures. In total, the experiment design was 3 x 4 x 6 (Texture x Size x Shape) with five repetitions for each cell to minimize the human error factor. For each drawing trial, the touch surface is tilted at a random angle within 0 to $90^{\circ}$ from the ground.

\subsection{Results}
There were above 72,000 data points collected in total. We measured the Euclidean error in 3D position and 3D orientation of the directional vector of the data points reported by 3DTouch and OptiTrack. The mean position error is 1.10 mm ($\sigma$ = 0.87), and the orientation error is 2.33 degrees ($\sigma$ = 2.58). The position error is a high overall error given the small target area sizes. As a relative reference, optical mice with similar resolution of 400 cpi, and frame rate of 1500 fps used in mobile robot odometry measurement had the maximum error below 0.8 mm in a 50 mm range \cite{palacin2006optical}.
\begin{figure}[!h]
\centering
\includegraphics[width=1.0\columnwidth]{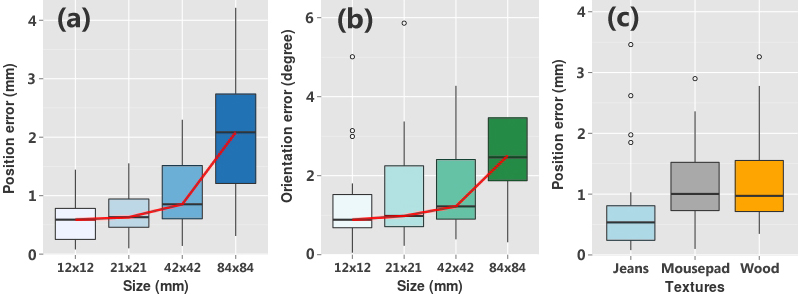}
\caption{(a) The mean position errors (mm) across 4 target sizes. (b) The mean orientation errors (degree) across 4 target sizes. (c) The mean position errors across 3 textures are not found to significantly differ.
}
\label{fig:figure_error}
\end{figure}

The results showed that the mean position and orientation errors increase with the target sizes (Figure~\ref{fig:figure_error}a, b). The three textures tested all have a high surface quality between 50-90. According to our Analysis of Variance (ANOVA) test, the position errors across the textures (Figure~\ref{fig:figure_error}c) did not show significant difference with \textit{F} = 2.227 and \textit{p} = 0.12.

We also realized a number of data clouds with high error generated by 3DTouch have largely variable distances from point to point. This suggests our position error is also partly due to the inherent acceleration error (i.e., up to 30g) in the ADNS-9800 sensor. This suggests that a more reliable optical sensor such as ADNS-2030 with 0.5g acceleration may significantly improve the performance. Hence, the results presented in this paper should be the baseline performance.

\section{Discussion and Future work}
Our accuracy evaluation showed that 3DTouch is capable of performing 3D translation with the mean error of 1.10 mm and 2.33 degrees within 84x84 mm target mobile touch area. However, a user study will be further conducted to measure usability feedback, especially fatigue and comfort level of our device. Also, we would like to further evaluate the performance of 3DTouch against the existing 3D input devices such as Wiimote, and mobile touch devices across different 3D virtual environment platforms (e.g., desktop, home theater, or CAVE).

3DTouch is a unique 3D input device that is designed with many useful potential use cases bringing 3D applications closer to users. First of all, 3DTouch is modular, allowing users to wear 2 pieces of the device, each on an index finger to enable multi-touch interaction. Moreover, the device has a flexible design, supporting multiple form factors. This allows users to wear the device at his comfort finger position, depending on specific interaction techniques. On the other hand, 3DTouch is wearable opening up a whole new design space for interaction techniques. We foresee the following several potential interaction techniques using 3DTouch for future work research: 
\begin{itemize}
\item In a CAVE, with 3DTouch worn on the index finger, users can use the palm of the other hand, or the thumb of the same hand as the touch surface.
\item Two or more fingers wearing pieces of 3DTouch would enable multi-touch interaction.
\item 3DTouch users can interact with curved surfaces as illustrated in Figure~\ref{fig:figure_translation}c. This allows users to interact with spherical and other non-flat displays.
\end{itemize}
 
\section{Conclusion}
In this paper, we present a novel 3D wearable input device using a combination of a laser optical sensor, and a 9-DOF inertial measurement unit. 3DTouch enables users to user their fingers or thumb as a 3D input device with the capability of performing 3D selection translation, and rotation. Our evaluation, shows the device's overall average tracking accuracy of 1.10 mm and 2.33 degrees within four target sizes from 12x12 mm to 84x84 mm, and across three different textures of jeans, mousepad, and wooden desk.

3DTouch is designed to fill the missing gap of a 3D input device that is self-contained, mobile, and universally working across various 3D platforms. This paper presents a low-cost solution to designing and implementing such a device. Modular solutions like 3DTouch opens up a whole new design space for interaction techniques to further develop on. With 3DTouch, we attempted to bring 3D interaction and applications a step closer to users in everyday life.

\section{Acknowledgments}

We would like to sincerely thank Dr. Jerry Hamann, George Janack, and Dr. Steven Barrett for their invaluable advice and help with the electrical aspects of this project. This work has been partly supported by the UW School of Energy fellowship.

%
%
%
%
%
\balance


\bibliographystyle{acm-sigchi}
\bibliography{sample}
\end{document}